\documentclass{svmult}
\usepackage{epsfig,graphicx} 
\usepackage[bottom]{footmisc}
\usepackage{natbib,url}      
\bibpunct{(}{)}{;}{a}{}{,}   
\def\cite#1{\citealp{#1}}    
\def\authorindex#1{}  
\def\figspath{.}  

\begin{document}\newcount\preprintheader\preprintheader=1


\def\thisvolume{these proceedings}

\def\aj{{AJ}}			
\def\araa{{ARA\&A}}		
\def\apj{{ApJ}}			
\def\apjl{{ApJ}}		
\def\apjs{{ApJS}}		
\def\ao{{Appl.\ Optics}} 
\def\apss{{Ap\&SS}}		
\def\aap{{A\&A}}		
\def\aapr{{A\&A~Rev.}}		
\def\aaps{{A\&AS}}		
\def\an{{Astron.\ Nachrichten}}
\def\aspcs{{ASP Conf.\ Ser.}}
\def\assp{{Astrophys.\ \& Space Sci.\ Procs., Springer, Heidelberg}}
\def\azh{{AZh}}			
\def\baas{{BAAS}}		
\def\jrasc{{JRASC}}	
\def\memras{{MmRAS}}		
\def\mnras{{MNRAS}}
\def\nat{{Nat}}		
\def\pra{{Phys.\ Rev.\ A}} 
\def\prb{{Phys.\ Rev.\ B}}		
\def\prc{{Phys.\ Rev.\ C}}		
\def\prd{{Phys.\ Rev.\ D}}		
\def\prl{{Phys.\ Rev.\ Lett.}} 
\def\pasp{{PASP}}
\def\pasj{{PASJ}}		
\def\qjras{{QJRAS}}
\def\science{{Sci}}		
\def\skytel{{S\&T}}		
\def\solphys{{Solar\ Phys.}} 
\def\sovast{{Soviet\ Ast.}}  
\def\ssr{{Space\ Sci.\ Rev.}}
\def\svassp{{Astrophys.\ Space Sci.\ Procs., Springer, Heidelberg}}
\def\zap{{ZAp}}			
\let\astap=\aap
\let\apjlett=\apjl
\let\apjsupp=\apjs
\def\grl{{Geophys.\ Res.\ Lett.}}  
\def\jgr{{J. Geophys.\ Res.}} 

\def\ion#1#2{{\rm #1}\,{\uppercase{#2}}}  
\def\deg{\hbox{$^\circ$}}
\def\sun{\hbox{$\odot$}}
\def\earth{\hbox{$\oplus$}}
\def\la{\mathrel{\hbox{\rlap{\hbox{\lower4pt\hbox{$\sim$}}}\hbox{$<$}}}}
\def\ga{\mathrel{\hbox{\rlap{\hbox{\lower4pt\hbox{$\sim$}}}\hbox{$>$}}}}
\def\sq{\hbox{\rlap{$\sqcap$}$\sqcup$}}
\def\arcmin{\hbox{$^\prime$}}
\def\arcsec{\hbox{$^{\prime\prime}$}}
\def\fd{\hbox{$.\!\!^{\rm d}$}}
\def\fh{\hbox{$.\!\!^{\rm h}$}}
\def\fm{\hbox{$.\!\!^{\rm m}$}}
\def\fs{\hbox{$.\!\!^{\rm s}$}}
\def\fdg{\hbox{$.\!\!^\circ$}}
\def\farcm{\hbox{$.\mkern-4mu^\prime$}}
\def\farcs{\hbox{$.\!\!^{\prime\prime}$}}
\def\fp{\hbox{$.\!\!^{\scriptscriptstyle\rm p}$}}
\def\micron{\hbox{$\mu$m}}
\def\onehalf{\hbox{$\,^1\!/_2$}}	
\def\onethird{\hbox{$\,^1\!/_3$}}
\def\twothirds{\hbox{$\,^2\!/_3$}}
\def\onequarter{\hbox{$\,^1\!/_4$}}
\def\threequarters{\hbox{$\,^3\!/_4$}}
\def\ubv{\hbox{$U\!BV$}}		
\def\ubvr{\hbox{$U\!BV\!R$}}		
\def\ubvri{\hbox{$U\!BV\!RI$}}		
\def\ubvrij{\hbox{$U\!BV\!RI\!J$}}		
\def\ubvrijh{\hbox{$U\!BV\!RI\!J\!H$}}		
\def\ubvrijhk{\hbox{$U\!BV\!RI\!J\!H\!K$}}		
\def\ub{\hbox{$U\!-\!B$}}		
\def\bv{\hbox{$B\!-\!V$}}		
\def\vr{\hbox{$V\!-\!R$}}		
\def\ur{\hbox{$U\!-\!R$}}


\def\labelitemi{{\bf --}}  

\def\rmit#1{{\it #1}}              
\def\rmit#1{{\rm #1}}              
\def\etal{\rmit{et al.}}           
\def\etc{\rmit{etc.}}           
\def\ie{\rmit{i.e.,}}              
\def\eg{\rmit{e.g.,}}              
\def\cf{cf.}                       
\def\viz{\rmit{viz.}}
\def\vs{\rmit{vs.}}

\def\rot{\hbox{\rm rot}}
\def\div{\hbox{\rm div}}
\def\lesssim{\mathrel{\hbox{\rlap{\hbox{\lower4pt\hbox{$\sim$}}}\hbox{$<$}}}}
\def\gtrsim{\mathrel{\hbox{\rlap{\hbox{\lower4pt\hbox{$\sim$}}}\hbox{$>$}}}}
\def\mathstacksym#1#2#3#4#5{\def#1{\mathrel{\hbox to 0pt{\lower 
    #5\hbox{#3}\hss} \raise #4\hbox{#2}}}}
\mathstacksym\lesssim{$<$}{$\sim$}{1.5pt}{3.5pt} 
\mathstacksym\gtrsim{$>$}{$\sim$}{1.5pt}{3.5pt} 
\mathstacksym\lrarrow{$\leftarrow$}{$\rightarrow$}{2pt}{1pt} 
\mathstacksym\lessgreat{$>$}{$<$}{3pt}{3pt} 

\def\dif{\: {\rm d}}                       
\def\ep{\:{\rm e}^}                        
\def\dash{\hbox{$\,-\,$}}                  
\def\is{\!=\!}                             

\def\starname#1#2{${#1}$\,{\rm {#2}}}  
\def\Teff{\hbox{$T_{\rm eff}$}}   

\def\kms{\hbox{km$\;$s$^{-1}$}}
\def\ms{\hbox{m$\;$s$^{-1}$}}
\def\Mxcm{\hbox{Mx\,cm$^{-2}$}}    

\def\Bapp{\hbox{$B_{\rm app}$}}    

\def\komega{($k, \omega$)}                 
\def\kf{($k_h,f$)}                         
\def\VminI{\hbox{$V\!\!-\!\!I$}}           
\def\IminI{\hbox{$I\!\!-\!\!I$}}           
\def\VminV{\hbox{$V\!\!-\!\!V$}}           
\def\Xt{\hbox{$X\!\!-\!t$}}                

\def\level #1 #2#3#4{$#1 \: ^{#2} \mbox{#3} ^{#4}$}   

\def\specchar#1{\uppercase{#1}}    
\def\AlI{\mbox{Al\,\specchar{i}}}  
\def\BI{\mbox{B\,\specchar{i}}} 
\def\BII{\mbox{B\,\specchar{ii}}}  
\def\BaI{\mbox{Ba\,\specchar{i}}}  
\def\BaII{\mbox{Ba\,\specchar{ii}}} 
\def\CI{\mbox{C\,\specchar{i}}} 
\def\CII{\mbox{C\,\specchar{ii}}} 
\def\CIII{\mbox{C\,\specchar{iii}}} 
\def\CIV{\mbox{C\,\specchar{iv}}} 
\def\CaI{\mbox{Ca\,\specchar{i}}} 
\def\CaII{\mbox{Ca\,\specchar{ii}}} 
\def\CaIII{\mbox{Ca\,\specchar{iii}}} 
\def\CoI{\mbox{Co\,\specchar{i}}} 
\def\CrI{\mbox{Cr\,\specchar{i}}} 
\def\CriI{\mbox{Cr\,\specchar{ii}}} 
\def\CsI{\mbox{Cs\,\specchar{i}}} 
\def\CsII{\mbox{Cs\,\specchar{ii}}} 
\def\CuI{\mbox{Cu\,\specchar{i}}} 
\def\FeI{\mbox{Fe\,\specchar{i}}} 
\def\FeII{\mbox{Fe\,\specchar{ii}}} 
\def\FeIX{\mbox{Fe\,\specchar{ix}}}
\def\FeX{\mbox{Fe\,\specchar{x}}}
\def\FeXVI{\mbox{Fe\,\specchar{xvi}}}
\def\FrI{\mbox{Fr\,\specchar{i}}}
\def\HI{\mbox{H\,\specchar{i}}} 
\def\HII{\mbox{H\,\specchar{ii}}} 
\def\Hmin{\hbox{\rmH$^{^{_{\scriptstyle -}}}$}}      
\def\Hemin{\hbox{{\rm He}$^{^{_{\scriptstyle -}}}$}} 
\def\HeI{\mbox{He\,\specchar{i}}} 
\def\HeII{\mbox{He\,\specchar{ii}}} 
\def\HeIII{\mbox{He\,\specchar{iii}}} 
\def\KI{\mbox{K\,\specchar{i}}} 
\def\KII{\mbox{K\,\specchar{ii}}} 
\def\KIII{\mbox{K\,\specchar{iii}}} 
\def\LiI{\mbox{Li\,\specchar{i}}} 
\def\LiII{\mbox{Li\,\specchar{ii}}} 
\def\LiIII{\mbox{Li\,\specchar{iii}}} 
\def\MgI{\mbox{Mg\,\specchar{i}}} 
\def\MgII{\mbox{Mg\,\specchar{ii}}} 
\def\MgIII{\mbox{Mg\,\specchar{iii}}} 
\def\MnI{\mbox{Mn\,\specchar{i}}} 
\def\NI{\mbox{N\,\specchar{i}}}
\def\NIV{\mbox{N\,\specchar{iv}}}
\def\NaI{\mbox{Na\,\specchar{i}}}
\def\NaII{\mbox{Na\,\specchar{ii}}}
\def\NaIII{\mbox{Na\,\specchar{iii}}}
\def\NeVIII{\mbox{Ne\,\specchar{viii}}} 
\def\NiI{\mbox{Ni\,\specchar{i}}} 
\def\NiII{\mbox{Ni\,\specchar{ii}}}
\def\NiIII{\mbox{Ni\,\specchar{iii}}} 
\def\OI{\mbox{O\,\specchar{i}}} 
\def\OVI{\mbox{O\,\specchar{vi}}}
\def\RbI{\mbox{Rb\,\specchar{i}}} 
\def\SII{\mbox{S\,\specchar{ii}}} 
\def\SiI{\mbox{Si\,\specchar{i}}} 
\def\SiII{\mbox{Si\,\specchar{ii}}} 
\def\SrI{\mbox{Sr\,\specchar{i}}}
\def\SrII{\mbox{Sr\,\specchar{ii}}}
\def\TiI{\mbox{Ti\,\specchar{i}}} 
\def\TiII{\mbox{Ti\,\specchar{ii}}} 
\def\TiIII{\mbox{Ti\,\specchar{iii}}} 
\def\TiIV{\mbox{Ti\,\specchar{iv}}} 
\def\VI{\mbox{V\,\specchar{i}}} 
\def\HtwoO{\mbox{H$_2$O}}        
\def\Otwo{\mbox{O$_2$}}          

\def\Halpha{\mbox{H\hspace{0.1ex}$\alpha$}} 
\def\Ha{\mbox{H\hspace{0.2ex}$\alpha$}}
\def\Hbeta{\mbox{H\hspace{0.2ex}$\beta$}}
\def\Hgamma{\mbox{H\hspace{0.2ex}$\gamma$}}
\def\Hdelta{\mbox{H\hspace{0.2ex}$\delta$}}
\def\Hepsilon{\mbox{H\hspace{0.2ex}$\epsilon$}}
\def\Hzeta{\mbox{H\hspace{0.2ex}$\zeta$}}
\def\Lyalpha{\mbox{Ly$\hspace{0.2ex}\alpha$}}
\def\Lybeta{\mbox{Ly$\hspace{0.2ex}\beta$}}
\def\Lygamma{\mbox{Ly$\hspace{0.2ex}\gamma$}}
\def\Lycont{\mbox{Ly\hspace{0.2ex}{\small cont}}}
\def\Baalpha{\mbox{Ba$\hspace{0.2ex}\alpha$}}
\def\Babeta{\mbox{Ba$\hspace{0.2ex}\beta$}}
\def\Bacont{\mbox{Ba\hspace{0.2ex}{\small cont}}}
\def\Paalpha{\mbox{Pa$\hspace{0.2ex}\alpha$}}
\def\Bralpha{\mbox{Br$\hspace{0.2ex}\alpha$}}

\def\NaD{\mbox{Na\,\specchar{i}\,D}}    
\def\NaDone{\mbox{Na\,\specchar{i}\,\,D$_1$}}
\def\NaDtwo{\mbox{Na\,\specchar{i}\,\,D$_2$}}
\def\NaID{\mbox{Na\,\specchar{i}\,\,D}}
\def\NaIDone{\mbox{Na\,\specchar{i}\,\,D$_1$}}
\def\NaIDtwo{\mbox{Na\,\specchar{i}\,\,D$_2$}}
\def\Done{\mbox{D$_1$}}
\def\Dtwo{\mbox{D$_2$}}

\def\Mgbone{\mbox{Mg\,\specchar{i}\,b$_1$}}
\def\Mgbtwo{\mbox{Mg\,\specchar{i}\,b$_2$}}
\def\Mgbthree{\mbox{Mg\,\specchar{i}\,b$_3$}}
\def\MgIb{\mbox{Mg\,\specchar{i}\,b}}
\def\MgIbone{\mbox{Mg\,\specchar{i}\,b$_1$}}
\def\MgIbtwo{\mbox{Mg\,\specchar{i}\,b$_2$}}
\def\MgIbthree{\mbox{Mg\,\specchar{i}\,b$_3$}}

\def\CaIIK{\mbox{Ca\,\specchar{ii}\,K}}       
\def\CaIIH{\mbox{Ca\,\specchar{ii}\,H}}
\def\CaIIHK{\mbox{Ca\,\specchar{ii}\,H\,\&\,K}}
\def\HK{\mbox{H\,\&\,K}}
\def\Kthree{\mbox{K$_3$}}      
\def\Hthree{\mbox{H$_3$}}
\def\Ktwo{\mbox{K$_2$}}
\def\Htwo{\mbox{H$_2$}}
\def\Kone{\mbox{K$_1$}}     
\def\Hone{\mbox{H$_1$}}     
\def\KtwoV{\mbox{K$_{2V}$}}
\def\KtwoR{\mbox{K$_{2R}$}}
\def\KoneV{\mbox{K$_{1V}$}}
\def\KoneR{\mbox{K$_{1R}$}}
\def\HtwoV{\mbox{H$_{2V}$}}
\def\HtwoR{\mbox{H$_{2R}$}}
\def\HoneV{\mbox{H$_{1V}$}}
\def\HoneR{\mbox{H$_{1R}$}}

\def\hk{\mbox{h\,\&\,k}}
\def\kthree{\mbox{k$_3$}}    
\def\hthree{\mbox{h$_3$}}
\def\ktwo{\mbox{k$_2$}}
\def\htwo{\mbox{h$_2$}}
\def\kone{\mbox{k$_1$}}     
\def\hone{\mbox{h$_1$}}     
\def\ktwoV{\mbox{k$_{2V}$}}
\def\ktwoR{\mbox{k$_{2R}$}}
\def\koneV{\mbox{k$_{1V}$}}
\def\koneR{\mbox{k$_{1R}$}}
\def\htwoV{\mbox{h$_{2V}$}}
\def\htwoR{\mbox{h$_{2R}$}}
\def\honeV{\mbox{h$_{1V}$}}
\def\honeR{\mbox{h$_{1R}$}}

\ifnum\preprintheader=1     
\makeatletter  
\def\@maketitle{\newpage
\markboth{}{}%
  {\mbox{} \vspace*{-8ex} \par 
   \em \footnotesize To appear in ``Magnetic Coupling between the Interior 
       and the Atmosphere of the Sun'', eds. S.~S.~Hasan and R.~J.~Rutten, 
       Astrophysics and Space Science Proceedings, Springer-Verlag, 
       Heidelberg, Berlin, 2009.} \vspace*{-5ex} \par
 \def\lastand{\ifnum\value{@inst}=2\relax
                 \unskip{} \andname\
              \else
                 \unskip \lastandname\
              \fi}%
 \def\and{\stepcounter{@auth}\relax
          \ifnum\value{@auth}=\value{@inst}%
             \lastand
          \else
             \unskip,
          \fi}%
  \raggedright
 {\Large \bfseries\boldmath
  \pretolerance=10000
  \let\\=\newline
  \raggedright
  \hyphenpenalty \@M
  \interlinepenalty \@M
  \if@numart
     \chap@hangfrom{}
  \else
     \chap@hangfrom{\thechapter\thechapterend\hskip\betweenumberspace}
  \fi
  \ignorespaces
  \@title \par}\vskip .8cm
\if!\@subtitle!\else {\large \bfseries\boldmath
  \vskip -.65cm
  \pretolerance=10000
  \@subtitle \par}\vskip .8cm\fi
 \setbox0=\vbox{\setcounter{@auth}{1}\def\and{\stepcounter{@auth}}%
 \def\thanks##1{}\@author}%
 \global\value{@inst}=\value{@auth}%
 \global\value{auco}=\value{@auth}%
 \setcounter{@auth}{1}%
{\lineskip .5em
\noindent\ignorespaces
\@author\vskip.35cm}
 {\small\institutename\par}
 \ifdim\pagetotal>157\p@
     \vskip 11\p@
 \else
     \@tempdima=168\p@\advance\@tempdima by-\pagetotal
     \vskip\@tempdima
 \fi
}
\makeatother     
\fi

\title*{Magnetic Coupling in the Quiet Solar Atmosphere}

\author{O. Steiner}

\authorindex{Steiner, O.} 

\institute{Kiepenheuer-Institut f\"ur Sonnenphysik, Freiburg, Germany}

\maketitle

\setcounter{footnote}{0}  

\begin{abstract} 
  Three kinds of magnetic couplings in the quiet solar atmosphere are
  highlighted and discussed, all fundamentally connected to the
  Lorentz force. First the coupling of the convecting and overshooting
  fluid in the surface layers of the Sun with the magnetic
  field. Here, the plasma motion provides the dominant force, which
  shapes the magnetic field and drives the surface dynamo. Progress in
  the understanding of the horizontal magnetic field is summarized and
  discussed. Second, the coupling between acoustic waves and the
  magnetic field, in particular the phenomenon of wave conversion and
  wave refraction.  It is described how measurements of wave travel
  times in the atmosphere can provide information about the topography
  of the wave conversion zone, i.e., the surface of equal Alfv\'en and
  sound speed.  In quiet regions, this surface separates a highly
  dynamic magnetic field with fast moving magnetosonic waves and
  shocks around and above it from the more slowly evolving field of
  high-beta plasma below it.  Third, the magnetic field also couples
  to the radiation field, which leads to radiative flux channeling and
  increased anisotropy in the radiation field. It is shown how faculae
  can be understood in terms of this effect. The article starts with
  an introduction to the magnetic field of the quiet Sun in the light
  of new results from the Hinode space observatory and with a brief
  survey of measurements of the turbulent magnetic field with the help
  of the Hanle effect.
\end{abstract}

\section{The magnetic field of the quiet Sun}      \label{steiner_sec_quiet}

Over the past three and a half years, the Sun stayed in a minimum
state of magnetic activity as it has ended cycle 23 and is about to
start with cycle 24 (if not pausing for yet a longer period of
time). In this period of quiescence it was possible to observe the Sun
with an exceptional instrument, the Solar Optical Telescope SOT
onboard the Hinode space observatory \citep{steiner_kosugi+al07}. The
Japanese Hinode satellite was put in orbit on September 22, 2006. It
is not so much the spatial resolution of 0.3\arcsec\ that makes this
instrument exceptional for quiet-Sun observing, but rather the absence
of seeing in combination with high pointing accuracy. This allows for
unprecedented ``deep'' (long-exposure) polarimetry with
correspondingly high polarimetric sensitivity at a spatial resolution
of 0.3\arcsec. A similar polarimetric accuracy at this high spatial
resolution has not been achieved from the ground in the past.

Immediately evident in the total or circular polarization maps over a
large field of view from Hinode \citep{steiner_lites+al08} is the
magnetic network, which persisted existing during the most quiet
states of the Sun. The magnetic network consists of an accumulation of
magnetic fields in the borders between supergranular cells. The origin
of these magnetic fields remains an enigma. Is it the result of
advection by the supergranular flow to which it is often described, or,
vice versa, is the supergranular flow rather a consequence of the
existence of the magnetic network (R.~Stein, private communication)?
Is the network field generated locally, near the
surface, or is it an integral part of the globally acting dynamo
\citep{steiner_stein+al03}, or is it just the decay product of sunspots
and/or ephemeral active regions?

Also omnipresent in this quiescent state of the Sun are small-scale
magnetic field concentrations, visible as delicate, bright objects 
within and at vortices of intergranular lanes. 
The structure made up of ensembles of bright elements 
is known as the \emph{filigree} \citep{steiner_dunn+zirker73}. 
\citet{steiner_mehltretter74}, while observing in the visible continuum,
referred to them as \emph{facular points\/} because they are
the footpoints of magnetic field concentrations that
appear as faculae near the solar limb. In more recent times, 
these objects were mostly observed in the G band 
(a technique originally introduced by \citealt{steiner_muller1985})
because the molecular band-head of CH that constitutes the G band, 
acts as a leverage for the intensity contrast 
\citep{steiner_rutten99,steiner_rutten+al01,
steiner_sanchez-almeida+al01,steiner_steiner+al01,steiner_shelyag+al04}.
Being located in the blue part of the visible spectrum, this 
choice also helps improving the diffraction-limited spatial resolution 
and the contrast in the continuum. Recent observational investigations of
the dynamics, morphology and properties of small-scale magnetic field 
concentrations of the quiet Sun include
\citet{steiner_berger+al04,steiner_langhans+al04,steiner_lites+socas-navarro04,
steiner_sanchez-almeida+al04,steiner_socas-navarro+lites04,
steiner_wiehr+al04,steiner_rouppe_van_der_voort+al05,
steiner_dominguez-cerdena+al06a,steiner_dominguez-cerdena+al06b,
steiner_berger+al07,steiner_bovelet+wiehr07,steiner_centeno+al07,
steiner_ishikawa+al07,steiner_langangen+al07,steiner_martinez-gonzalez+al07,
steiner_orozco-suarez+al07,
steiner_rezaei+al07a,steiner_rezaei+al07b,steiner_tritschler+al07,
steiner_bello-gonzalez+kneer08a,
steiner_bello-gonzalez+al08b,steiner_de_wijn+al08,
steiner_orozco-suarez+al08,steiner_bello-gonzalez+al09}
and references therein. It would be interesting to learn to
which degree the abundance of small-scale magnetic field concentrations 
persists in a grand minimum, to find out more about their origin.

With the help of the spectropolarimeter of the 
Solar Optical Telescope (SOT) onboard Hinode 
it became for the fist time possible to reliably 
determine the transversal (with respect to the line-of-sight)
magnetic field component of the quiet Sun. These measurements
indicate that, seen with a spatial resolution of 0.3\arcsec, 
the quiet internetwork regions harbor a photospheric magnetic 
field whose mean field strength of its horizontal component 
considerably surpasses that  of the vertical component  
\citep{steiner_lites+al07,steiner_lites+al08,steiner_orozco-suarez+al07}. 
According to 
these papers, the vertical fields are concentrated in the intergranular 
lanes, whereas the horizontal fields  occur most commonly at 
the edges of the bright granules, aside from the vertical fields. 
\citet{steiner_lites+al08} determine for the horizontal field component
a mean apparent field strength (averaged over a large field of view 
including network and internetwork regions) of 55\,G, while the 
corresponding mean absolute vertical field strength was only 11\,G. 
\citet{steiner_harvey+al07} find from recordings with GONG and SOLIS 
at moderate angular resolution a ``seething magnetic field'' with a 
line-of-sight component increasing from disk 
center to limb as expected for a nearly horizontal field orientation.
It is reasonable to assume that the horizontal fields of
\citet{steiner_lites+al08} and those of \citet{steiner_harvey+al07}
are different manifestations of the same magnetic field.
\citet{steiner_ishikawa+al08} detected transient horizontal magnetic fields
in plage regions as well. Previously, \citet{steiner_lites+al96}  and
\citet{steiner_meunier+al98} reported observations of weak and strong
horizontal fields in quiet Sun regions. 

The anisotropy of the quiet-Sun magnetic field as revealed by the Hinode 
measurements is not necessarily in contradiction with
the observed depolarization of scattered light through the Hanle 
effect. For the quantitative interpretation of the Hanle effect it was 
customarily assumed in the past that the quiet Sun magnetic field was 
``turbulent'' in the sense that it was isotropically distributed on
the scales relevant for the analysis.
Theories have now been developed to include distribution functions 
for the field strength and angular directions
(see, e.g., \cite{steiner_carroll+kopf07}, \cite{steiner_sampoorna+al08},
\cite{steiner_sampoorna09}) for a better description of the turbulent
magnetic field in radiation transfer and first steps are taken to use 
such formulations for the computation of the Hanle effect in scattering 
media \citep{steiner_frisch06,steiner_nagendra+al09}.

The isotropy assumption may still be valid on scales smaller than
0.3\arcsec. In fact, simulations suggest constant angular distribution
within $\pm 50\deg$ from the horizontal direction on a scale of
$0.05\arcsec$. On the other hand, should the field be strongly
anisotropic on all scales, it would still produce Hanle depolarization
but its interpretation would be less straightforward. From a
theoretical point of view, the anisotropy of the magnetic field comes
not as a surprise, and homogeneous turbulence of the magnetic field
seems unlikely since the convective flow is far from homogeneously
turbulent given the scale size of 1000\,km of granules vs.\ the
pressure scale height of 100\,km.

\citet{steiner_stenflo1982} roughly estimated the strength of the weak 
turbulent magnetic field based on the Hanle technique between 10 and 100\,G.
This range was narrowed to 4--40\,G by \citet{steiner_stenflo+al98}.
More precise estimates by \citet{steiner_faurobert93}  and
\citet{steiner_faurobert+al95} yielded values in the range of 
30 to 60\,G in the deep photosphere and 10 to 20\,G in the middle 
and upper photosphere, where the latter values were increased to
20 to 30\,G by \citet{steiner_faurobert+al01}. 
The values reported by Stenflo et al.\ and by Faurobert et al.\ are
not without controversy, however.
\citet{steiner_trujillo-bueno+al04} find from three-dimensional 
radiative transfer modelling of scattering polarization in atomic 
and molecular lines an ubiquitous tangled magnetic field with an 
average strength of 130\,G  around 300\,km height in the photosphere, 
which is much stronger in the intergranular than in the granular regions.
They estimate that the energy density of this field would amount
to 20\% of the kinetic energy density of the convective motion at 
a height of 200\,km. If this high value is correct
it also indicates that the Zeeman measurements with Hinode
\citep{steiner_lites+al07,steiner_lites+al08,steiner_orozco-suarez+al07}
have not captured quite all of the existing quiet-Sun fields, presumably 
because of polarimetric cancellation which Zeeman measurements
are subject to in contrast to Hanle measurements. 
Adopting the idealized model of a single-valued microturbulent field,
\citet{steiner_trujillo-bueno+al04} obtained a mean field strength
that varies between 50 to 70\,G in the height range from 400\,km to
200\,km, respectively, and only when taking an exponential probability
density function for the field-strength distribution into account do
they obtain the higher value of 130\,G. Relatively high values are
also reported by \citet{steiner_bommier+al05}.

More in line with Faurobert et al.\ are \citet{steiner_shapiro+al07},
who obtain from differential Hanle measurements with the CN violet
system a field strength in the range from 10 to 30\,G in the upper
solar photosphere, while the analyses of the observed scattering
polarization in C2 lines by
\citet{steiner_faurobert+arnaud03} and \citet{steiner_berdyugina+fluri04}
imply a field strength of about 10\,G. However measurements of the
scattering polarization in molecular lines may be quite sensitive to
the thermal structure in the atmosphere.  In fact
\citet{steiner_trujillo-bueno+al04} obtain from measurements with
C$_2$ a field strength of the order of 10\,G too, but they also show
that these measurements sample the atmosphere mainly above granules
only, where, correspondingly, the turbulent field must be much weaker
than in the downflows of the intergranular space (see the review by
\cite{steiner_trujillo-bueno+al06} for a detailed presentation of their 
Hanle-effect measurements).

In any case, it seems that ``deep'' polarimetric measurements with
Hinode have discovered a large part of the hitherto ``hidden''
magnetic field that was known to us only through Hanle
measurements. It made this field accessible to Zeeman analysis and
therefore to a more reliable determination of its angular
distribution, at least down to $0.3\arcsec$ spatial resolution.  In
the next chapter, we review results from recent simulation that aim at
explaining the predominance of the mean horizontal over the mean
vertical field in the quiet-Sun photosphere.

\section{Coupling of convection with magnetic fields}    \label{steiner_sec_convec}

It was mentioned in the previous chapter that a few observational
studies prior to Hinode already hinted at a frequent occurrence of
horizontally oriented magnetic fields in the quiet Sun. Likewise, the
horizontal fields did not come unannounced to theoretical solar physics.
\citet{steiner_ugd+al98} noted
``we find in all simulations also strong horizontal fields above
convective upflows'', and \citet{steiner_schaffenberger+al05,
steiner_schaffenberger+al06} found frequent horizontal fields in their
three-dimensional simulations, which they describe as ``small-scale
canopies''.  Also the three-dimensional simulations of
\citet{steiner_abbett07} display ``horizontally directed ribbons of
magnetic flux that permeate the model chromosphere'', not unlike the
figures shown by
\citet{steiner_schaffenberger+al06}.
However, these reports did not receive wide attention because 
actual measurement of the weak transversal component was not possible
or unreliable prior to the advent of Hinode. 

\begin{figure}  
  \centering
  \includegraphics[width=0.43\textwidth]{\figspath/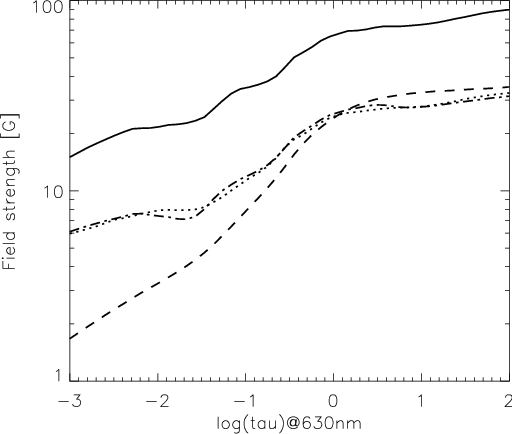}
  \hfill
  \includegraphics[width=0.48\textwidth]{\figspath/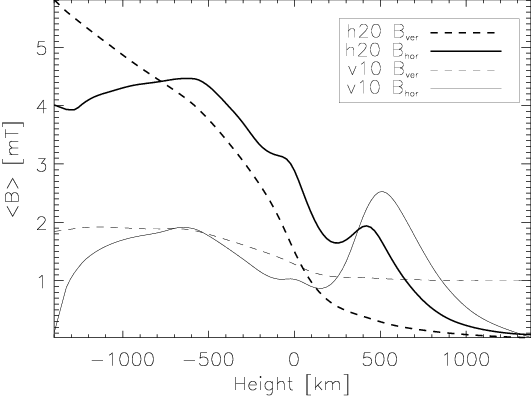}   
  \caption[]{\label{steiner_fig01}
  {\em Left\/}: Mean absolute horizontal magnetic field components, 
  $\langle |B_{x}|\rangle$ ($-\cdot-\cdot-\cdot-$), and 
  $\langle |B_{y}|\rangle$ ($\ldots\ldots$), and absolute
  vertical field component, $\langle |B_{\mathrm{ver}}|\rangle$ 
  (-\,-\,-\,-\,-) as a function of optical depth 
  $\log\tau_{\,630\,{\rm nm}}$ of the dynamo run of 
  \citet{steiner_schuessler+voegler08}. The averaging refers to surfaces of
  constant $\tau_{\,630\,{\rm nm}}$. The solid curve is the rms
  ${\langle B_{x}^2 + B_{y}^2\rangle}^{1/2}$. {\em Right\/}:
  $\langle B_{\mathrm{hor}}\rangle$ = 
  $\langle (B_{x}^2 + B_{y}^2)^{1/2}\rangle$ (---------)
  and $\langle |B_{\mathrm{ver}}|\rangle$ (-- -- -- -- --) as
  a function of height $z$ from the simulation run h20 
  (heavy) and run v10 (thin) of \citet{steiner_steiner+al08}. v10
  and h20 substantially differ in their initial and boundary 
  conditions for the magnetic field. Note the different
  physical meanings of the abscissa and the different units
  and scales in the ordinates of the two plots.   
}\end{figure}

More recently and after the discovery of the horizontal field with
Hinode, two theoretical works (\cite{steiner_schuessler+voegler08} and
\cite{steiner_steiner+al08}) specifically aimed at finding out more
about its nature and origin. Both papers present results of
three-dimensional magnetohydrodynamic numerical simulations of the
internetwork magnetic field with regard to the intrinsically produced
horizontal magnetic field. In the following I briefly summarize and
compare part of their results.

The two simulation runs presented by \citet{steiner_steiner+al08} and
the ``local dynamo run'' of \citet{steiner_schuessler+voegler08} 
differ substantially in their initial and boundary conditions for the
magnetic field. Yet, they all show a clear dominance of the horizontal
field in parts or the full height range where the spectral lines used
for the Hinode observations are formed. Thus, the intrinsic production
of a predominantly horizontal magnetic field in the photosphere of
three-dimensional magnetohydrodynamic simulations is a rather robust
result. Figure~\ref{steiner_fig01} shows the horizontal and the
vertical magnetic field strengths as a function of height in the
atmosphere of the three simulations. Left and right boundaries of the
left panel correspond to approximately $z= 400$\,km and $z =
-1000$\,km of the right panel, respectively. Note that the scale of
the ordinate is logarithmic and in gauss in the left panel but linear
and in mT in the right panel.  Also account for the non-linear
relation between the abscissa of the two panels. Interestingly, both
simulation runs of
\citet{steiner_steiner+al08} show a local maximum of the horizontal field 
component near 500\,km height and this is also the case for a 
local dynamo run when the top (open) boundary is located at $z=650$\,km 
(M.~Sch\"ussler private communication).

How do these results compare with Hinode? For a fair comparison 
it is indispensable to synthesize the Stokes profiles of the 
630\,nm \ion{Fe}{i} spectral line pair from the simulations
and subsequently derive whatever parameters were derived from 
the actual observations. The analysis of the synthetic data must 
proceed in the very same manner as done with the observed profiles. 
Applying the appropriate point spread function 
\citep{steiner_wedemeyer08} to the synthetic profiles and subjecting 
them to the same procedure for conversion to apparent flux densities 
as done by \citet{steiner_lites+al08} for the observed profiles, 
\citet{steiner_steiner+al08} obtain spatial and temporal averages for 
the transversal and longitudinal apparent magnetic flux densities,
$|B^{\mathrm{T}}_{\mathrm{app}}|$ and $|B^{\mathrm{L}}_{\mathrm{app}}|$  
of respectively 21.5\,G and 5.0\,G for run h20 and 
10.4\,G and 6.6\,G for run v10.
Thus, the ratio 
$r=\langle |B^{\mathrm{T}}_{\mathrm{app}}|\rangle/
   \langle |B^{\mathrm{L}}_{\mathrm{app}}|\rangle = 4.3$ 
for h20 and $1.6$ for v10. \citet{steiner_lites+al08} obtain
from Hinode SP data 
$\langle |B^{\mathrm{T}}_{\mathrm{app}}|\rangle = 55$\,G and 
$\langle |B^{\mathrm{L}}_{\mathrm{app}}|\rangle = 11$\,G 
resulting in $r = 5.0$. Run v10 was judged to rather reflect network fields 
because of  its preference to produce vertically directed, unipolar 
magnetic fields, enforced by its initial and boundary conditions.
For the internetwork field, h20 is more appropriate. Correspondingly,
the $r$-value of h20 better agrees with the measurements of
\citet{steiner_lites+al08}, which measures mainly internetwork magnetic
fields. It should be cautioned that $r$ is quite dependent on spatial
resolution in the sense that lower resolution overestimates this value.
The reason for this behavior is that horizontal fields have a more patchy,
smoother, and less intermittent character than the vertical fields and
are therefore less subject to polarimetric cancellation.\footnote{Remember
that polarimetric cancellation also occurs for independent field 
components of the horizontal field when they are perpendicular to each 
other within a single pixel area. Polarimetric cancellation
occurs for horizontal fields in the same way as for vertical fields --
transversal perpendicular fields lead to polarimetric cancellation  
just as antiparallel longitudinal fields do.} For a comparison between
synthetic and observed center-to-limb data see \citet{steiner_steiner+al09}.

\begin{figure}  
  \centering
  \includegraphics[width=1.0\textwidth]{\figspath/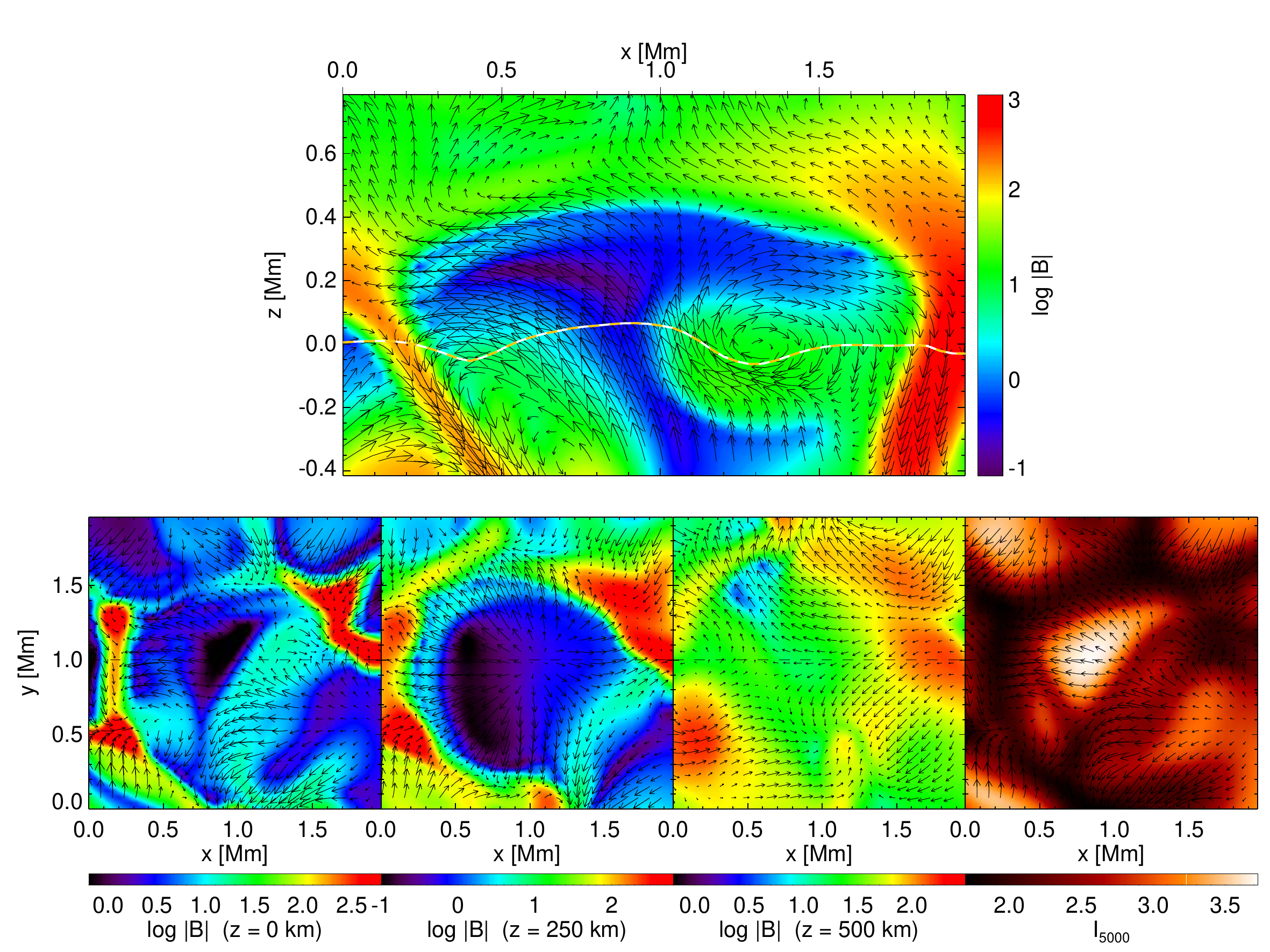}
  \caption[]{\label{steiner_fig02} Flux expulsion in a close-up from a
  MHD simulation by \citet{steiner_schaffenberger+al05}: Logarithmic
  magnetic field strength in a vertical cross-section (top) and in
  three horizontal cross-sections (bottom) at heights of 0\,km,
  250\,km, and 500\,km. The emergent intensity is displayed in the
  rightmost panel. The arrows represent the velocity field in the
  shown projection planes. The white curve in the upper panel marks
  the height of optical depth unity. From
  \citet{steiner_wedemeyer+al_issi08}.  }\end{figure}

What kind of physical process produces the horizontal fields? 
\citet{steiner_schuessler+voegler08} and \citet{steiner_steiner+al08}
offer two different but not necessarily exclusive explanations. Rather 
they emphasize two different aspects of the coupling of convection with 
magnetic fields that is at the origin of the horizontal fields.
\citet{steiner_steiner+al08} emphasize the aspect of the flux
expulsion process \citep{steiner_weiss1966,steiner_galloway+weiss81},
which describes the expulsion of magnetic field from the interior
of an eddy flow like that of granules. Thus, the fact that the magnetic
field tends to be located in the intergranular space and not
within granules is considered a consequence of the flux expulsion process.
However it should be noted that the granular flow is not bounded
alone by intergranular lanes but also by the overlaying photosphere,
which efficiently damps overshooting flow owing to its superadiabatic 
stratification. Hence, magnetic field tends not only to be expelled
in the lateral direction to the intergranular lanes but also in the 
vertical direction, where it accumulates in the upper photosphere and 
lower chromosphere. 
In fact, vertical sections through the computational domain such as
Fig.~1 of \citet{steiner_schuessler+voegler08} and Fig.~3 of
\citet{steiner_schaffenberger+al05} show magnetic voids 
where the granular flow is most vigorous as a consequence of the flux 
expulsion process. The voids are arched by horizontally directed 
magnetic field. This can be nicely seen in the close-up shown in 
Fig.~\ref{steiner_fig02}.

\begin{figure}  
  \centering
  \includegraphics[width=1.0\textwidth]{\figspath/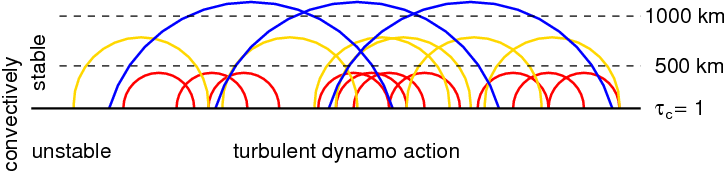}
  \caption[]{\label{steiner_fig03}
  Dynamo action amplifies and maintains the magnetic field in the
  convectively unstable layer below $\tau_{\mathrm{c}} = 1$. In the 
  layers above, the field is mainly determined by its distribution at 
  the $\tau_{\mathrm{c}} = 1$ surface. This configuration leads to a 
  steep decline of the absolute vertical flux with height as can be 
  seen by counting the loop foot-points at each indicated level because 
  small loops are more abundant than large loops. On the other hand, 
  this configuration leads to a less steep decline of the mean horizontal 
  field strength.
}\end{figure}

\citet{steiner_schuessler+voegler08} emphasize the aspect from the
local dynamo that operates in the convectively unstable layers
beneath the surface of continu\-um optical depth unity. Near the
surface, weak magnetic field gets quickly stretched and thus amplified
by the convective flow, in particular also by the small-scale turbulent 
flow of intergranular downflows.\footnote{There is nothing mystic 
about this amplification, which is a natural consequence of the field 
being tied to the plasma in (quasi) ideal MHD, which does work against 
the Lorentz force on the expense of kinetic energy. However, 
\citet{steiner_voegler+schuessler07} were able to demonstrate that a 
local dynamo operates in these layers, which means that
a magnetic field of constant mean energy density is maintained without  
the need of continuous supply of a weak (seed) field. It even survives 
when downflows continuously pump magnetic field out of the simulation box.} 
On the other hand, in the convectively stable photosphere above, the flows 
become weaker and field amplification rapidly drops with height. Thus,
the magnetic field in the photosphere and its decay with height 
is mainly determined by the field distribution at the surface 
$\tau_{\mathrm{c}} = 1$, in particularly by its energy spectrum as a function 
of horizontal wave number, which in turn is determined by the turbulent 
dynamo beneath this surface. This results in a steep decline of the absolute 
vertical magnetic flux with height as can be seen from 
Fig.~\ref{steiner_fig03}. While many loops of small scales (where the
energy spectrum is maximal) contribute to the vertical flux in the deep 
photosphere, fewer loops of large scales (where the energy is less)
add to it in the higher layers. On the other hand, this configuration leads
to a less steep decline of the horizontal field and hence, the
mean horizontal field strength starts to dominate the mean absolute 
vertical field strength as a function of height. In this picture
the dominance of the horizontal field component is a natural outcome 
of the dynamo-generated field.

\begin{figure}  
  \centering
  \includegraphics[width=0.5\textwidth]{\figspath/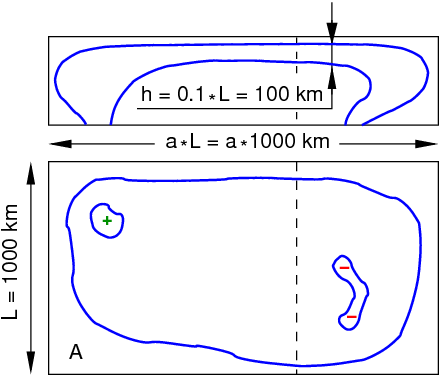}
  \caption[]{\label{steiner_fig04}
  Two intergranular magnetic flux concentrations of opposite polarity 
  are connected by a small-scale loop (canopy) in the photosphere,
  spanning a distance of granular scale. Even though this magnetic
  field configuration is divergence free, it follows that the measured
  mean absolute horizontal field is much stronger than the mean absolute 
  vertical field:
  $\langle |B_{\rm h}|\rangle/\langle |B_{\rm v}|\rangle \approx 5a$.
}\end{figure}

It was argued in the course of this conference by J.O.~Stenflo
that a predominance of the horizontal field
over the vertical one was in contradiction with the solenoidality
condition for the magnetic field. Leaving aside that the simulations
strictly maintain solenoidality and still show a predominance of the 
horizontal over the vertical component, Fig.~\ref{steiner_fig04} 
provides another counter example to this conjecture. Assume that
within an area $A$  of granular size $L$ there are two vertical flux 
concentrations of, say $B_{\rm v}500$\,G, that occupy an area of 
$f_{\rm v} A$, where $f_{\rm v} \approx 0.02$. 
The two flux concentrations of opposite
polarity are connected by a photospheric, solenoidal arch of 
thickness $h\approx 0.1 L$ as it occurs in simulations (viz.\ the 
``small-scale canopies'' of \cite{steiner_schaffenberger+al05}).
Then flux conservation demands that  in a cross section of the loop 
(as the one indicated by the dashed line)
$\Phi = B_{\rm h} L h = B_{\rm v} f_{\rm v} A/2$, where $B_{\rm v}$
is the vertical field strength at $\tau_{\mathrm{c}} \approx 1$ 
and $B_{\rm h}$ the strength of the horizontal field of the arch.
It follows that $B_{\rm h} = 5 a f_{\rm v} B_{\rm v}$. With
$a\approx 1$ we obtain $B_{\rm h}\approx 50$\,G. When the horizontal
field of the arch fills about $f_h = 0.8$ of the area $A$, we obtain
a mean horizontal field of 
$\langle |B_{\rm h}|\rangle = B_{\rm h} f_h = 
  B_{\rm v} f_{\rm h} f_{\rm v} 5 a = 40$\,G while
$\langle |B_{\rm v}|\rangle = B_{\rm v} f_v = 10$\,G.
The ratio
$\langle |B_{\rm h}|\rangle/\langle |B_{\rm v}|\rangle = 5 a f_{\rm h}$
can be made arbitrarily large by increasing $a$, i.e., by
stretching the arch.

Confusion may arise because of misinterpreting
transversal Zeeman measurements as being a measure of magnetic 
flux. It is true that for longitudinal Zeeman measurements, the 
measured mean flux density is directly proportional to the 
magnetic flux (disregarding polarimetric cancellation effects for now). 
Such a generalization is not valid for the transversal Zeeman effect,  
where the mean flux density is an average value over an area
that is not perpendicular but parallel to the measured field 
component. In this case the apparent flux density is a
spatially averaged flux density, not more. Assuming that the 
horizontal field has the vertical extent of the atmospheric 
scale height, $h$, with a filling factor of 1, then the horizontal 
magnetic flux is $\Phi_{\rm h} = \langle B_{\rm h}\rangle h L$, 
where $L$ is the scale of the field of view. The vertical flux is
$\Phi_{\rm v} = \langle B_{\rm v}\rangle L^2$ and consequently, 
using flux conservation
$\langle B_{\rm h}\rangle/\langle B_{\rm v}\rangle \propto L/h$. 
On granular scales
$L/h \approx 1000\,{\rm km}/100\,{\rm km} = 10$, which again 
suggests that the predominance of the apparent horizontal flux
density over the vertical one is a direct consequence of the anisotropic
nature of convective turbulence in the highly stratified atmosphere
of the Sun, where the granular scale is ten times larger than the
pressure scale height. However, remember that this calculation does not 
apply to subgranular scales at which the field may still be
isotropic.


\section{Coupling of waves with magnetic fields}          \label{steiner_sec_waves}

\begin{figure}
\centering
\includegraphics[width=0.48\textwidth]{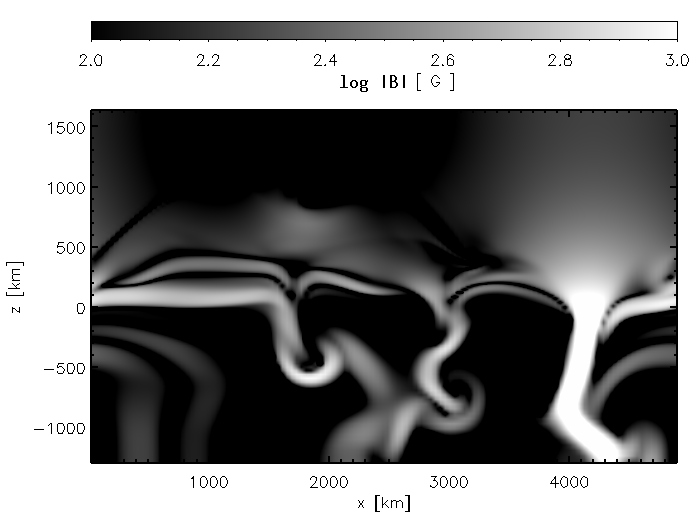}
\includegraphics[width=0.48\textwidth]{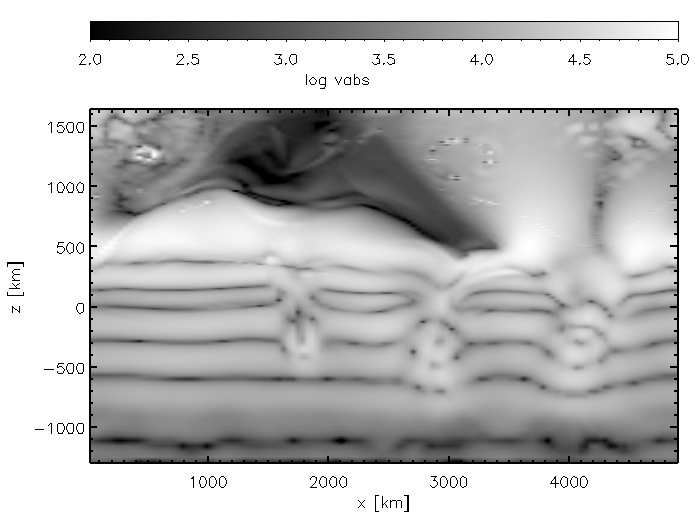}
\caption[]{\label{steiner_fig05}
  {\em Left\/}: Logarithmic absolute magnetic flux density in a
  two-dimensional simulation domain. Magnetic flux concentrations form 
  in the downdrafts of convection. A particularly strong one has formed 
  near $x = 4150$\,km. {\em Right\/}: 
  A plane-parallel wave of frequency 20\,mHz travels through the convecting 
  plasma into the magneti\-cally structured photosphere and further into the
  low-$\beta$ (magnetically dominated) chromosphere. The panel
  shows the difference in absolute velocity between the perturbed and the
  unperturbed solution 168\,s after launching the wave. The magnetic field 
  at launch time corresponds to that of the panel to the left.
  Optical depth $\tau_{\,500\,{\rm nm}} = 1$ is close to $z=0$.
  The velocity scaling is logarithmic with dimension [cm/s].
  At the location of the magnetic flux concentration the initially fast (acoustic)
  wave has converted character to fast magnetic and it underwent refraction
  to such a degree that the wave front extending from $(x,z) = (2400,1500)$ 
  to $(3400,500)$ has already completely turned around and travels back 
  {\it into} the atmosphere  again. A similar fanning out of the wave front 
  starts to occur around $x=1100$\,km. Adapted from \citet{steiner_steiner+al07} 
  courtesy of Ch.~Nutto.
}\end{figure}

Acoustic waves are generated and emitted by the convectively
unstable layers beneath the solar surface ($\tau_{\,500\,{\rm nm}} = 1$).
They couple to the magnetic field when they enter the atmospheric layers 
above. Such coupling can be seen to take place in Fig.~\ref{steiner_fig05}, 
which shows a time instant of a two-dimensional simulation. Starting from a state where a
magnetic flux concentration has formed  at approximately $x = 4150$\,km
(left panel), a plane-parallel, monochromatic acoustic wave is introduced
at the bottom of the computational domain (right panel), which propagates 
within a time span of about 200\,s across the non-stationary atmosphere -- from 
the convection-zone, $z\in [-1200, 0]$\,km, through the photosphere, 
$z\in [0, 500]$\,km, into the magnetically dominated chromosphere, 
$z\in [500,1600]$\,km, where it gets partially refracted and reflected 
by interaction with the magnetic field. 

The perturbation of the wave front in the convection zone in 
Fig.~\ref{steiner_fig05} is not due to the presence of a magnetic 
field, but rather to the vigurous intergranular downflows and associated 
temperature deficit. However, as soon as 
the wave front enters the magnetically dominated atmosphere where 
$\beta\le 1$, coupling with the magnetic field kicks in and part of 
the wave gets primarily magnetically driven. ($\beta$ is the ratio of
magnetic to thermal pressure.)  The major effects of this 
interaction are that (i) the wave front speeds up since the Alfv\'en
velocity becomes the characteristic speed, which sharply increases 
with height when magnetic pressure drops less quickly than
the gas pressure, and (ii) the waves refract because of the 
inhomogeneous magnetic field, defining an inhomogeneous refractive 
index for the magneto-acoustic wave.

When measuring the wave travel time between two fixed geometrical height levels in the 
atmosphere (representing the formation height of two spectral lines) we observe a
decrease at locations of strong magnetic field concentrations because of (i). 
Moreover, due to (ii) the wave behaves like evanescent because of the strong refraction 
that effectively leads to a reflection of the wave. Therefore, the wave travel time 
betrays the presence of the magnetic field concentration and it can be used to map the 
topography of the magnetic field in the solar atmosphere. In fact, this effect was 
employed by \citet{steiner_finsterle+al04} in order to obtain the three-dimensional 
topography of the ``magnetic canopy'' in and around active regions by determining the 
travel time of high-frequency acoustic waves in the solar chromosphere.
\citet{steiner_steiner+al07}, from where Fig.~\ref{steiner_fig05} is derived from,
demonstrated with the help of numerical experiments that wave travel-times can indeed 
serve this purpose.

The theory and theoretical aspects of magneto-acoustic waves in a
gravitationally stratified atmosphere, sometimes called
magneto-acoustic-gravity (MAG) waves or magneto-atmospheric waves,
have received much attention in recent years.  Pioneering works
include those of \citet{steiner_thomas1982} and
\citet{steiner_zhugzhda+dzhalilov1982,
steiner_zhugzhda+dzhalilov1984a, steiner_zhugzhda+dzhalilov1984b,
steiner_zhugzhda+dzhalilov1984c}.  \citet{steiner_shibata1983} carried
out initial numerical computer experiments with magneto-atmospheric
waves. In more recent times,
\citet{steiner_rosenthal+al02} and \citet{steiner_bogdan+al03} published two 
comprehensive papers on the subject. These works include several numerical 
experiments with non-uniform magnetic field equilibria in a two-dimensional,
stratified atmosphere. They recognized and highlighted  the role of 
refraction of fast magnetic waves and the role of the surface of equal 
Alfv\'en and sound speed as a wave conversion zone. Aiming at applications 
in local helio-seismology,
\citet{steiner_cally05} derives gravito-magneto-acoustic 
dispersion relations and then uses these to examine how acoustic rays 
entering regions of strong field split into fast and slow components
and the subsequent fates of each. \citet{steiner_cally07} presents the 
theory in a particularly instructive manner. Results from
numerical simulations of MAG-wave propagation in three-dimensional
space are presented by \citet{steiner_cally+goossens08} and
\citet{steiner_moradi+al09}.

\citet{steiner_khomenko+collados06} carried out numerical simulations of 
magne\-to-acoustic wave propagation in sunspots and found that the fast 
(magnetic) mode in the region where $c_{\rm s} < v_{\rm A}$ does not reach the 
chromosphere but reflects back to the photosphere due to wave refraction, 
caused primarily by the vertical and horizontal gradients of the Alfv\'en 
speed. For small-scale flux-tubes,
\citet{steiner_khomenko+al08b} find that deep horizontal motions
of the flux tube initially generate a slow (magnetic) mode and a
surface mode that are efficiently transformed into a slow (acoustic)
mode when the magnetic field starts to dominate. This slow mode
propagates vertically along the magnetic field remaining always within
the flux tube, where it steepens to a shock. Only a small part of the
driver energy is returned to the photosphere by the fast
magneto-acoustic mode. \citet{steiner_khomenko+al08a} demonstrate that
photospheric five-minute oscillations can leak into the chromosphere
inside small-scale vertical magnetic flux tubes as a consequence of
radiative damping, which leads to a significant reduction of the
cutoff frequency and they provide observational evidences of this
effect. This effect is not to be confounded with the ``ramp effect''
\citep{steiner_cally07}, which lowers the cutoff frequency when the
flux tube is inclined with respect to the gravitational acceleration
\citep{steiner_suematsu1990,steiner_jefferies+al06}.  Both these works
of Khomenko et al.\ suggest that vertical magnetic field
concentrations play an essential role in coupling the dynamics of the
photosphere to the chromosphere through efficient channeling and
conversion of magneto-acoustic waves.

\begin{figure}
\centering
\includegraphics[width=0.32\textwidth]{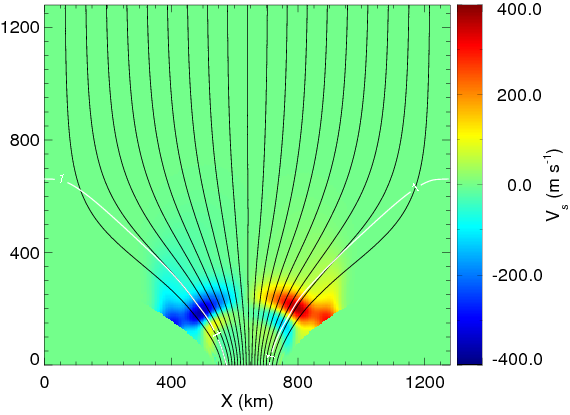}
\includegraphics[width=0.32\textwidth]{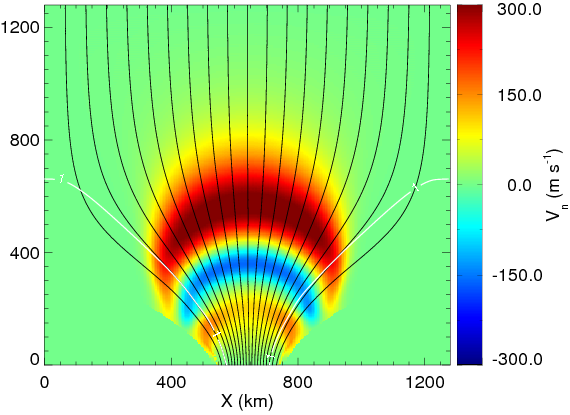}
\includegraphics[width=0.32\textwidth]{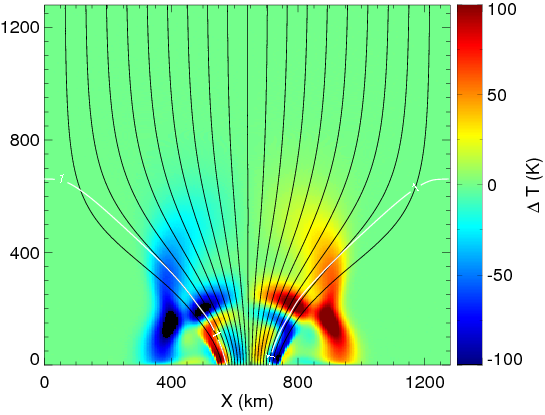}
\caption[]{\label{steiner_fig06}
Snapshot of magneto-acoustic waves, generated by an initial impulsive
movement of the equilibrium magnetic flux concentration (black field lines) 
to the right. {\em Left\/}: Longitudinal velocity of the slow 
(predominantly acoustic) wave that propagates along the magnetic field. 
{\em Middle\/}: Transversal velocity due to the fast (predominantly magnetic)
wave within the flux sheet. It undergoes differential refraction. {\em Right\/}:
Temperature perturbation due to these two waves and due to the fast acoustic wave
in the ambient medium. The white contours indicates $\beta = 1$. 
Velocities are only shown where $|B|>50$\,G.
Courtesy G.~Vigeesh.
}\end{figure}

Figure~\ref{steiner_fig06} demonstrates the complexity of magneto-acoustic wave 
propagation in a magnetically structured, stratified atmosphere. A magnetic
flux sheet (two-dimensional) of a strength of 1600\,G at its base (where 
$\beta < 1$) is shifted to the right in the transverse direction with a single 
impulse of 12\,s duration and a maximal velocity of 0.75\,km/s. As a consequence 
of this sudden movement, four types of waves emanate from the base of the 
flux sheet: (i) a fast (magnetic) wave in the low-$\beta$ regime of the flux-sheet
interior, visible in the middle panel, (ii) a slow (acoustic) wave that 
propagates along the magnetic field within the flux sheet, visible in the
left panel, (iii) a fast acoustic wave that propagates spherically
into the ambient medium, visible in the right panel, and (iv) a slow magnetic
wave that propagates in the boundary layer of the flux sheet.  The predominantly 
acoustic waves show an antisymmetric pattern with respect to the sheet axis because 
of the movement to the right, which causes a compression at the leading edge 
and a decompression at the trailing edge of the flux sheet. However, as the 
slow wave propagates it becomes asymmetric (not visible in Fig.~\ref{steiner_fig06}). 
On the left side the wave crest quickly steepens into a shock because of being 
preceded by a wave trough. On the right side the inverse sequence causes the 
wave to spread, which impedes at first the development of a shock. 

The shape of the fast (magnetic) wave becomes
crescent (middle panel) because of the non-uniform Alfv\'en speed. As the tips
of the crescents move essentially sideways, they enter the low-$\beta$ periphery
of the flux sheet where they convert to predominantly acoustic (fast) and therefore
become visible in $\delta T$ (right panel) as the wing like feature that extends 
in the vertical direction. From a rough estimate of the acoustic energy flux
generated by such impulsive transverse motions, \citet{steiner_vigeesh+al09}
conclude that this flux would hardly balance the chromospheric energy 
requirements in the network. Previously, similar numerical experiments 
with impulsive and periodic driving have been carried out by 
\citet{steiner_hasan+al05} and \citet{steiner_hasan+ballegooijen08}.

Until here we have considered the coupling of waves with intense, small-scale
magnetic flux concentrations like they occur in network and plage 
regions.  The numerical experiments mentioned above, all make use of 
highly idealized wave drivers: typically a monochromatic transversal or 
longitudinal periodic motion, or a single impulsive motion is imposed on 
an initially static equilibrium configuration.
Certainly for the internetwork, such quasi static states are unrealistic. 
Furthermore, magnetic flux concentrations in the intergranular lanes of the 
internetwork attain typically hectogauss not kilogauss field strength and 
they rather connect with the omnipresent horizontal field than 
vertically extending into the chromosphere. In this case, the strength of 
the internetwork field can be expected to exponentially decrease with height 
like the gas pressure and the density do, so that the magnetic field would
not become dominant in the upper layers of the photosphere and the 
chromosphere and consequently no substantial coupling between waves and the 
magnetic field would occur in these layers. 
On the other hand if there is a predominance of one magnetic 
polarity, part of the magnetic flux can be expected to connect
to the outer solar atmosphere or to a region of opposite polarity further 
away, in which case the field strength would decrease less steeply leading
to $\beta \le 1$ in the chromosphere (but not yet in the photosphere). 
How do waves couple to the magnetic field under these circumstances?

\begin{figure}
\centering
\includegraphics[width=0.30\textwidth]{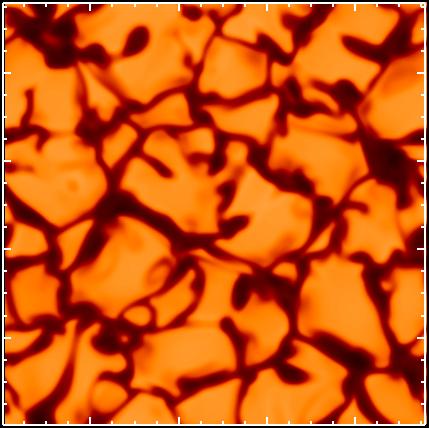}
\hspace*{0.03\textwidth}
\includegraphics[width=0.30\textwidth]{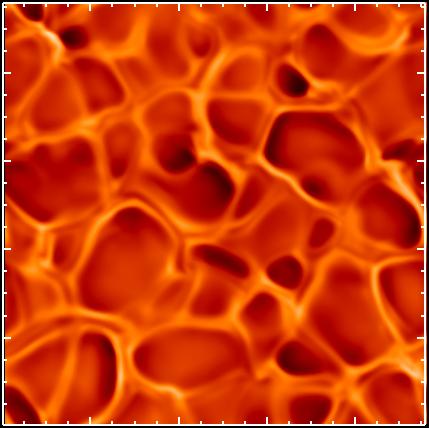}
\hspace*{0.03\textwidth}
\includegraphics[width=0.30\textwidth]{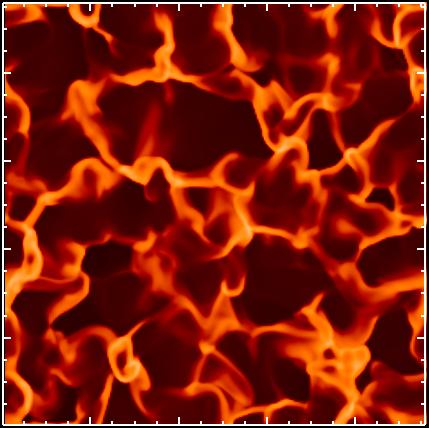}
\caption[]{\label{steiner_fig07}
Three horizontal cross sections through a a simulation domain with 9.6\,Mm
side length displaying the temperature. {\em Left\/}: Section at $z = -200$\,km
showing granules (hot) and intergranular lanes (cool). {\em Middle\/}: Section
at $z = +200$\,km showing inverse granulation. {\em Right\/}: Section at
$z = +1000$\,km showing the ``fluctosphere'' consisting of a rapidly
changing network pattern of hot material compressed in traveling shock waves. 
Pockets of cool, expanded material reside in between the hot plasma.
}\end{figure}

First to the waves. Figure~\ref{steiner_fig07} shows the temperature in
three horizontal cross sections through the computational domain of 9.6\,Mm
side length. At 200\,km below the $\tau_{\,500\,{\rm nm}} = 1$ surface 
we see the cool intergranular lanes and hot granules (left panel),
at $+200$\,km the hot intergranular lanes and cool granules of the 
inverse granulation (seen a bit higher up in spectroscopic
quantities). Both these patterns evolve on roughly the granular time
scale. In the cross section at $+1000$\,km we see a totally different pattern 
that evolves on a much shorter time scale and is due to shock waves 
that have formed at this height range and travel in all directions, 
forming a network of hot material. This shock-wave pattern that
emerges from acoustic waves, which are generated by the convective 
granular motion at the base of the atmosphere was first shown
to exist in three-dimensional simulations without magnetic field by 
\citet{steiner_wedemeyer+al04}. 
It leads to large fluctuations in the
tenuous atmosphere above the classical temperature minimum,
to a veritable ``fluctosphere'' \citep{steiner_wedemeyer+al_issi08},
earlier termed the ``clapotisphere'' by
\citet{steiner_rutten+uitenbroek1991}
and \citet{steiner_rutten1995} with regard to peak fluctuations caused
by the interference of (shock-)waves.



In combination with magnetic fields, these disturbances give rise to a
rich variety of magneto-acoustic wave phenomena. As detailed above,
the gas pressure in the gravitationally stratified atmosphere may drop
more quickly with height than the magnetic energy density does, giving
rise to a height range where sound speed and Alfv\'en speed are of
similar magnitude.  Within this region, which forms a corrugated
surface excursive over a wide height range in the three-dimensional
atmosphere, propagating wave modes change nature from acoustic to
magnetic and from slow to fast and vice versa. Above this surface
there is a predominant tendency for magnetic modes to get refracted
and reflected due to the dispersive nature of the inhomogeneous
magnetic field.

\citet{steiner_schaffenberger+al05} have simulated this case with 
a field of a predominant polarity of constant mean net vertical flux 
density 10\,G. Their three-dimensional simulation domain encompasses
a height range from $-1500$\,km to $+1500$\,km (where zero corresponds to
$\tau_{\,500\,{\rm nm}} = 1$). Immediately apparent from a movie that shows 
the field strength
(\url{http://www.kis.uni- freiburg.de/~steiner/vsec.mov}) 
is that the surface of $\beta = 1$ (where $c_{\rm s}\approx v_{\rm A}$)
separates a region of highly dynamic 
magnetic fields with fast moving magnetosonic waves and shocks around and 
above it from the more slowly evolving field of high-beta plasma below it.
This surface is located at roughly 1000\,km in this case.
It is corrugated and its local height strongly varies in time over a 
range of about 1000\,km.

The magnetic field in the chromosphere of this simulation
continuously rearranges
itself on a time scale of less than 1\,min, much shorter  than in the 
photosphere or in the convection-zone layers. The field has a 
strength between 2 and 40 G. Different from the surface magnetic field, 
it is more homogeneous and fills practically the entire space so that 
the magnetic filling factor in the top layer is close to unity. 
There seems to be no spatial correlation between chromospheric flux 
accumulations and the small-scale field concentrations in the photosphere. 
Magnetoacoustic waves that form transient filaments of stronger 
than average magnetic field are a ubiquitous phenomenon in the 
chromosphere. They form in the compression zone downstream and 
along propagating shock fronts. These magnetic filaments that 
have a field strength rarely exceeding 40\,G, rapidly move with 
the shock fronts and quickly form and dissolve with them.
Hence, the coupling of waves with the magnetic field leads to a continuous
agitation of the magnetic field in the chromosphere by shock waves.
It is not yet clear what the significance and the consequences of 
these perturbations are, especially in view of electro-magnetic dissipation 
processes.

\section{Coupling of radiation with magnetic fields}        \label{steiner_sec_rad}

\begin{figure}
\centering
\includegraphics[width=0.5\textwidth]{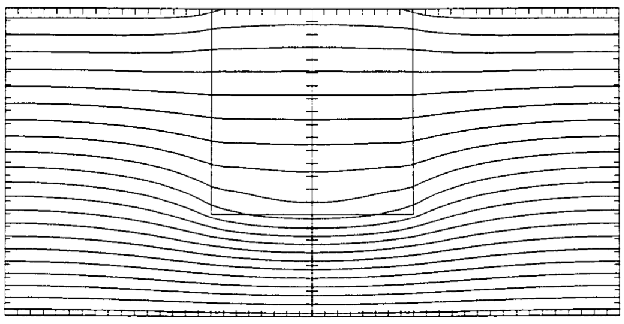}
\caption[]{\label{steiner_fig08}
Temperature contours of a two-dimensional schematic magnetic flux sheet
in radiative equilibrium with its surroundings. The opacity in the center 
rectangle is one fifth of the opacity in the rest of the domain, causing 
a redirection (channeling) of the radiative flux. In this state of radiative 
equilibrium the flux channeling causes a lowering of the temperature in the 
bottom part of the ``flux sheet'' and a temperature enhancement in the top part,
where radiation starts escaping to the ambient free space. 
Adapted from \citet{steiner_steiner1991}.
}
\end{figure}

The radiative flux in the solar atmosphere couples to the magnetic field
not only microscopically (changing polarization state) but also macroscopically 
through modification
of the gas pressure and density by the magnetic field. The magnetic field 
has an energy density, $e_{\rm mag}$, often called the magnetic pressure, 
like gas pressure has: $p_{\rm g} = e_{\rm therm}$. The ratio of the 
two is $\beta = e_{\rm therm}/e_{\rm mag}$. In locations where $\beta\le 1$, 
magnetic pressure substitutes gas pressure in the transversal direction, 
which at the same time lowers the material density (given thermodynamic 
equilibrium) and with it the opacity. Hence, as a rule of thumb, opacity is 
lower where the magnetic field is strong, 
which leads to a redirection of the radiative flux. This effect was
called ``flux channeling'' by \citet{steiner_cannon85}, because the
net radiative flux, which is strictly vertical and outwardly directed
in a plane parallel stellar atmosphere, becomes partially and
laterally redirected in the presence of a strong, vertically directed
magnetic flux concentration.  
In radiative equilibrium this
effect leads to a temperature perturbation in the sense that the
temperature is slightly enhanced in the surface layers of the magnetic
flux concentration but lowered in the deep layers as is shown in
Fig.~\ref{steiner_fig08}.


The coupling of the radiation field to the magnetic field via the 
channeling effect introduces a substantial anisotropy in the radiation field 
at the location of a magnetic flux concentration and in its
surroundings. This anisotropy gives rise to the facular phenomenon.

Faculae can be explained in terms of the ``hot wall model''
\citep{steiner_spruit76}. In this model the hot wall corresponds to the
wall of the depression that is caused by a magnetic flux
concentration, like the Wilson depression of a sunspot. The wall
separates the magnetic flux concentration from the surrounding,
practically field-free plasma. However, the term ``hot wall'' may be
misleading as it is by far not as hot as the plasma in the same height
range in the unperturbed ambient atmosphere. In fact, the rectangular
wall of the low-opacity region of the radiative equilibrium model
shown in Fig.~\ref{steiner_fig08} has in most parts a lower
temperature.  In the case of photospheric magnetic flux concentrations
the temperature of the wall is determined by a delicate balance
between radiative losses and the convective energy supply, where the
latter is actually reduced in the transverse direction close to the
flux concentration because the magnetic field acts like a solid
wall to plasma motion in this direction.  Equilibrium models and
magnetohydrodynamic simulations, however, invariably show a bright
edge where the wall of the ``Wilson depression'', merges with the
horizontal surface
\citep{steiner_carlsson+al04,steiner_keller+al04,steiner_steiner05,
       steiner_depontieu+al06}.
Seen at an oblique angle (corresponding to an observation off disk center) 
this ``hot edge'' leads to a characteristic center-to-limb variation in contrast. 

\begin{figure}
\sidecaption
\includegraphics[width=0.4\linewidth]{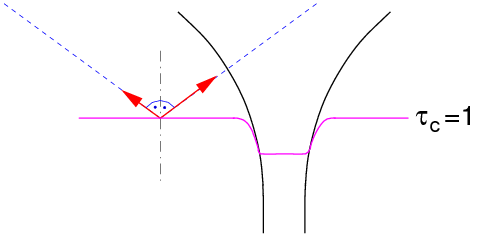}
\caption[]{\label{steiner_fig09}
Photons preferentially escape along the line of sight that
traverses the magnetic flux concentration because of its rarified (less opaque) 
atmosphere. Hence, the radiation field lateral to the flux concentration is
asymmetric due to radiative flux channeling.
}\end{figure}

It was only recently realized when analyzing high resolution filtergrams
of faculae from the Swedish Solar Telescope (SST) that their contrast enhancement 
extends quite a distance in the limbward direction, typically for half a granular 
size, which is much further than the depth of the depression wall would possibly 
be \citep{steiner_lites+al04,steiner_hirzberger+wiehr05}.
\citet{steiner_berger+al07} measure an average radial width of faculae of 400\,km.
This means that the contrast enhancement of faculae extends beyond the
depression proper in the limbward direction. The reason for this behavior is 
explained with the help of Fig.~\ref{steiner_fig09} as follows.

A material parcel located in the solar atmosphere lateral to a magnetic flux 
concentration ``sees'' a more  transparent atmosphere in the direction toward 
the flux concentration as compared to a direction under equal zenith angle but 
pointing away from it because of its rarefied atmosphere. 
Consequently, from a wide area surrounding the magnetic flux concentration, 
radiation escapes more easily in the direction  towards the flux sheet so that 
a single flux concentration impacts the radiative escape in a cross-sectional 
area that is much wider than the magnetic field  concentration proper. This
means that the presence of a magnetic flux concentration introduces
an anisotropy in the radiation field such that when observed close to the limb,
a granule limbward of (seen across) the flux concentration appears brighter than 
normal. Hence, a facula is not to be identified with 
bright plasma that sticks,  as the name may insinuate, like a torch out of 
the solar surface. Rather is it the manifestation of photospheric granulation, seen 
across a magnetic flux concentration -- granulation that appears brighter 
than normal in the form of so called ``facular granules''.




Another consequence of the flux channeling effect is an enhancement of
radiative loss from the solar surface at the location of the
small-scale magnetic flux concentration. One could say that magnetic
flux concentrations add a roughness to the solar surface that
increases its effective area and thus increases the radiative loss
from it. Therefore, the coupling between radiation and magnetic field,
and with it faculae, play a key role in the solar irradiance variation
over a solar cycle and on shorter time scales
\citep{steiner_fligge+al00,steiner_wenzler+al05,steiner_foukal+al06,
steiner_krivova+solanki08}.

\section{Conclusion}                   \label{steiner_sec_conclusion}

Results obtained with the spectropolarimeter of the 
Solar Optical Telescope onboard the Hinode space observatory  have
greatly extended our knowledge of the magnetism of the quiet
Sun. It has now become apparent that 
virtually every location on the surface of the Sun harbors magnetic
field that can be detected in the 630\,nm \ion{Fe}{i} spectral line
via the Zeeman effect, exceeding a $1\sigma$ noise level of 0.6\,G 
and 20\,G for the longitudinal and transversal magnetic field, 
respectively \citep{steiner_lites+al08}.
At the scale of $0.3\arcsec$, this magnetic 
field has a preferential direction, which is parallel to the solar 
surface, i.e., it is anisotropic. It seems now that Hinode 
has discovered the majority of the hitherto ``hidden'' magnetic 
field that was known to us only through Hanle measurements, and it made 
it accessible to Zeeman analysis.

The predominantly horizontal direction of the weak internetwork
magnetic field is also a robust result of magnetohydrodynamic 
simulations of the surface layers of the Sun.
It can be explained in terms of the coupling 
between magnetic field and convective plasma motion. Two aspects of 
this interaction are evident. First, the process of flux expulsion,
which displaces magnetic field from eddy cells. It leads to a
concentration of predominantly vertical fields in the interganular 
lanes and predominantly horizontal fields in the stable layers above
and at the edges immediately adjacent to granules. Second, the turbulent 
dynamo, which operates in the top surface layers of the convection 
zone, leads in the adjacent,
stably stratified photospheric layer to a multi-scale system of magnetic 
loops. This loop system naturally leads to a stronger decline with 
height of the vertical than of the horizontal component of the
magnetic field, hence, to a predominance of the horizontal fields
in the height range where the spectral lines used for 
the Hinode observations are formed. It is still a matter of future
research to find out to which degree the internetwork magnetic
field is due to the turbulent surface dynamo, the remnants of 
pre-existing magnetic flux of active regions, and the emergence 
of magnetic flux from the deep convection zone, or due to yet
other, additional sources. Also the role of the horizontal
field in the heating of the chromosphere, e.g., by Ohmic dissipation
of associated current sheets, needs yet to be clarified.

The study of the propagation of magnetoacoustic waves in a 
magnetically structured, stratified atmosphere is a relatively 
new field of research. It has applications in magnetoatmospheric
seismology, for example, when determining the topography of the
magnetic canopy by measuring wave travel times in the photosphere
and chromosphere. The surface defined by equality between sound
speed and Alfv\'en speed is a zone of mode conversion.
It separates the magnetically dominated tenuous 
region of fast moving magnetic modes and magnetic shock waves 
from the more slowly evolving atmosphere
beneath it. The strong gradients in the Alfv\'en speed of
a magnetically structured atmosphere lead to refraction and
reflection of the magnetically driven modes. 
Future research in this field is directed at improving diagnostics
for magnetoatmospheric seismology. Measurement and theory of
magnetoatmospheric wave propagation need to be improved and
linked for a reliable interpretation and exploitation
of observations. The role of wave mode conversion and the channeling
of slow modes in magnetic flux concentrations for the heating of
the outer atmosphere must yet be quantified. 

Magnetic flux concentrations lead to radiative flux channeling and 
increased anisotropy in the radiation field. Faculae can be understood
in terms of this effect, which is apparent from magnetohydrodynamic 
simulations in two and three spatial dimensions. Future research in
this field should include a thorough statistical analysis of (synthetic) 
faculae of three-dimensional simulations in order to understand the
role of magnetic field stength, size, shape, etc., for the center-to-limb
behavior of the facular contrast and for the energy balance of faculae.
The latter is crucial for a better understanding of solar radiance 
variability.

\begin{acknowledgement}
I am grateful to Ch.~Nutto, S.~Wedemeyer-B\"ohm, and G.~Vigeesh for help
with the figures, M.~Sch\"ussler and J.~Trujillo Bueno for clarifications
regarding the turbulent dynamo and Hanle depolarization, and Ch.~Bethge,
J.~Bruls, R.~Hammer, and Ch.~Nutto for proofreading individual sections.
This work was supported by German Academic Exchange Service (DAAD)
grant D/05/57687.
\end{acknowledgement}

\begin{small}


\bibliographystyle{rr-assp}       

\end{small}

\end{document}